\documentclass[twocolumn,a4paper,reprint,amssymb,aps,prd,groupedaddress,superscriptaddress,preprintnumbers,amsmath,nobibnotes,showpacs,nofootinbib]{revtex4-2}
\usepackage{graphicx}
\usepackage{epsf}
\usepackage{bm}

\usepackage{amsmath}
\usepackage[T1]{fontenc}
\usepackage{amsfonts}
\usepackage{amssymb}
\usepackage{epstopdf}
\usepackage{natbib}
\usepackage{color}
\usepackage[dvipsnames]{xcolor}
\usepackage{verbatim}
\usepackage{multirow}
\usepackage{physics}
\usepackage{pifont}
\usepackage{comment}
\usepackage{bm}
\usepackage{microtype}
\usepackage[colorlinks=true, linkcolor=blue , citecolor=blue, urlcolor=blue]{hyperref}
\usepackage{xcolor}
\usepackage{amsmath}
\usepackage[capitalize]{cleveref}
\usepackage[normalem]{ulem}
\usepackage{enumitem}
\usepackage{lipsum}
\usepackage{epstopdf}
\usepackage{etoolbox}

\usepackage{siunitx}        
\newcommand{\LCDM}{\ensuremath{\Lambda\mathrm{CDM}}}
\newcommand{\LsCDM}{\ensuremath{\Lambda_{\mathrm{s}}\mathrm{CDM}}}

\newcommand{\gLCDM}{\ensuremath{\gamma\Lambda\mathrm{CDM}}}
\newcommand{\gLsCDM}{\ensuremath{\gamma\Lambda_{\mathrm{s}}\mathrm{CDM}}}
\newcommand{\Ls}{\ensuremath{\Lambda_{\mathrm{s}}}}
\newcommand{\kunit}{\ensuremath{h\,\mathrm{Mpc}^{-1}}}

\newcommand{\be}[1]{\begin{equation}\label{#1}}
\newcommand{\ee}{\end{equation}}
\newcommand{\ba}[1]{\begin{eqnarray}\label{#1}}
\newcommand{\ea}{\end{eqnarray}}

\begin{document}

\title{Nonlinear Matter Power Spectrum from relativistic $N$-body Simulations: $\Lambda_{\rm s}$CDM versus $\Lambda$CDM}

\author{\"{O}zg\"{u}r Akarsu}
\email{akarsuo@itu.edu.tr}
\affiliation{Department of Physics, Istanbul Technical University, Maslak 34469 Istanbul, T\"{u}rkiye}

\author{Eleonora Di Valentino}
\email{e.divalentino@sheffield.ac.uk}
\affiliation{School of Mathematical and Physical Sciences, University of Sheffield, Hounsfield Road, Sheffield S3 7RH, United Kingdom}

\author{Ji\v{r}\'i Vysko\v{c}il}
\email{j.vyskocil@hzdr.de}
\affiliation{Center for Advanced Systems Understanding, Untermarkt 20, 02826 G\"{o}rlitz, Germany}
\affiliation{Helmholtz-Zentrum Dresden-Rossendorf, Bautzner Landstra\ss e 400, 01328 Dresden, Germany}

\author{Ezgi Y{\i}lmaz}
\email{ezgi.yilmaz@dzastro.de}
\affiliation{Deutsches Zentrum für Astrophysik, Postplatz 1, 02826, Görlitz, Germany}

\author{A. Emrah Y\"{u}kselci} 
\email{yukselcia@itu.edu.tr}
\affiliation{Department of Physics, Istanbul Technical University, Maslak 34469 Istanbul, T\"{u}rkiye}

\author{Alexander Zhuk}
\email{ai.zhuk2@gmail.com}
\affiliation{Center for Advanced Systems Understanding, Untermarkt 20, 02826 G\"{o}rlitz, Germany}
\affiliation{Helmholtz-Zentrum Dresden-Rossendorf, Bautzner Landstra\ss e 400, 01328 Dresden, Germany}
\affiliation{Astronomical Observatory, Odesa I.I. Mechnikov National University, Dvoryanskaya Street 2, Odesa 65082, Ukraine} 

\begin{abstract}
We present relativistic $N$-body simulations of a $\Lambda_{\rm s}$CDM---sign-switching cosmological constant (CC)---scenario under general relativity and compare its nonlinear matter power spectrum to $\Lambda$CDM at ${z = 15,\,2,\,1,\,0}$, using best-fit parameters from \emph{Planck}-only and a combined ``full'' dataset. During the AdS-like CC ($\Lambda_{\rm s}<0$) phase, prior to the transition redshift $z_\dagger$, reduced Hubble friction dynamically enhances the growth of perturbations; after the switch, with dS-like CC ($\Lambda_{\rm s}>0$), the larger late-time expansion rate partly suppresses, but does not erase, the earlier amplification. Consequently, the ratio $P_{\Lambda_{\rm s}\rm CDM}/P_{\Lambda\rm CDM}$ exhibits a pronounced, redshift-dependent shape feature: a crest peaking at ${\sim 20-25\%}$ around ${k \simeq 1-3\,h\,\mathrm{Mpc}^{-1}}$ near the transition, which then migrates to larger physical scales and persists to ${z = 0}$ as a robust ${\sim 15-20\%}$ uplift at ${k \simeq 0.6-1.0\,h\,\mathrm{Mpc}^{-1}}$. These wavenumbers correspond to group or poor-cluster environments and lie within the sensitivity range of weak lensing, galaxy-galaxy lensing, cluster counts, and tSZ power, providing a concrete, falsifiable target that cannot be mimicked by a scale-independent change in $\sigma_8$ or $S_8$. The timing (earlier for \emph{Planck}-only, later for the full dataset) and the amplitude of the crest align with the ``cosmic noon'' epoch (${z \simeq 1-2}$), offering a gravitational prior for the observed peak in the cosmic star-formation rate.
\end{abstract}


\maketitle

\section{Introduction}
\label{sec:intro}

The concordance \LCDM\ model remains the simplest and most successful framework for describing the evolution of the Universe, providing an excellent fit to early-universe probes such as the Cosmic Microwave Background (CMB) anisotropies~\cite{Planck:2018vyg, ACT:2020frw, SPT-3G:2025bzu} and Big Bang Nucleosynthesis (BBN). Nevertheless, in recent years a number of persistent \emph{cosmological tensions} have emerged when \LCDM\ predictions are confronted with late-time observations~\cite{Abdalla:2022yfr,Perivolaropoulos:2021jda,DiValentino:2022fjm,Akarsu:2024qiq,CosmoVerseNetwork:2025alb}. The most prominent is the more than $6\sigma$ discrepancy between the value of the Hubble constant ($H_0$)~\cite{Verde:2019ivm,DiValentino:2020zio,DiValentino:2021izs,Schoneberg:2021qvd,Shah:2021onj,Kamionkowski:2022pkx,Giare:2023xoc,Hu:2023jqc,Verde:2023lmm,DiValentino:2024yew,Perivolaropoulos:2024yxv} inferred from CMB data~\cite{Planck:2018vyg, ACT:2020frw, SPT-3G:2025bzu} and that measured by the local distance ladder~\cite{Freedman:2020dne,Birrer:2020tax,Anderson:2023aga,Scolnic:2023mrv,Jones:2022mvo,Anand:2021sum,Freedman:2021ahq,Uddin:2023iob,Huang:2023frr,Li:2024yoe,Pesce:2020xfe,Kourkchi:2020iyz,Schombert:2020pxm,Blakeslee:2021rqi,deJaeger:2022lit,Murakami:2023xuy,Breuval:2024lsv,Freedman:2024eph,Riess:2024vfa,Vogl:2024bum,Scolnic:2024hbh,Said:2024pwm,Boubel:2024cqw,Scolnic:2024oth,Li:2025ife,Jensen:2025aai,Riess:2025chq,Newman:2025gwg,Stiskalek:2025ktq,H0DN:2025lyy}, but additional anomalies, such as the so-called $S_8$ tension~\cite{DiValentino:2020vvd,DiValentino:2018gcu,Nunes:2021ipq,DES:2021bvc,DES:2021vln,KiDS:2020suj,Asgari:2019fkq,Joudaki:2019pmv,DAmico:2019fhj,Kilo-DegreeSurvey:2023gfr,Troster:2019ean,Heymans:2020gsg,Dalal:2023olq,Chen:2024vvk,ACT:2024okh,DES:2024oud,Harnois-Deraps:2024ucb,Dvornik:2022xap,DES:2021wwk,Wright:2025xka,Adil:2023jtu,Akarsu:2024hsu} in weak lensing reinforce the case that \LCDM, while highly successful, may be incomplete. 
Proposed solutions are commonly grouped into two broad categories, by when they act on the expansion history: \emph{early-time modifications}, which alter the expansion or energy content before recombination (e.g., early dark energy, EDE~\cite{Poulin:2018cxd,Karwal:2016vyq,Hill:2020osr,Kamionkowski:2022pkx,Ivanov:2020ril,Sakstein:2019fmf,Niedermann:2019olb,Niedermann:2020dwg,Poulin:2023lkg,Smith:2025grk,Poulin:2025nfb,SPT-3G:2025vyw}), and \emph{late-time modifications}, which deform the post-recombination expansion while preserving high-redshift successes of the standard cosmology (e.g., interacting dark energy, IDE~\cite{DiValentino:2017iww,DiValentino:2019ffd,Yang:2018euj,Yang:2018uae,Nunes:2016dlj,Giare:2024smz,Caprini:2016qxs,Yang:2021hxg,Yang:2017ccc,Pan:2020zza,Escamilla:2023shf,Bernui:2023byc,Gao:2021xnk,Li:2024qso,Costa:2018aoy,DiValentino:2020kpf,vonMarttens:2018iav,Silva:2025hxw,Yang:2025uyv,vanderWesthuizen:2025rip}). 
In this context, recent results from DESI~\cite{DESI:2024mwx,DESI:2025zgx} have sharpened the observational picture, showing that extensions such as \textit{dynamical dark energy} (DDE)~\cite{Giare:2024gpk,Gialamas:2024lyw,RoyChoudhury:2024wri,Dinda:2024kjf,Giare:2024oil,RoyChoudhury:2025dhe, RoyChoudhury:xxx, Scherer:2025esj,Pang:2025lvh,Roy:2024kni,Ormondroyd:2025iaf,Li:2025cxn,Cortes:2024lgw,Najafi:2024qzm,Wang:2024dka,Giare:2024ocw,Giare:2025pzu,Kessler:2025kju,Pang:2025lvh,Teixeira:2025czm,Specogna:2025guo,Sabogal:2025jbo,Cheng:2025lod,Herold:2025hkb,Cheng:2025hug,Ozulker:2025ehg,Lee:2025pzo,Silva:2025twg,Fazzari:2025lzd} can substantially improve the joint consistency of BAO and supernova data relative to \LCDM, further motivating systematic explorations of late-time departures from \LCDM\ and possible new physics affecting late-universe dynamics.

Among the realizations of the latter category, the \LsCDM\ framework (also known as the \textit{sign-switching cosmological constant} (CC))~\cite{Akarsu:2019hmw,Akarsu:2021fol,Akarsu:2022typ,Akarsu:2023mfb} stands out as one of the most promising and economical extensions of \LCDM. It posits that around redshift $z_\dagger\!\sim\!2$, the Universe underwent a rapid \textit{mirror} AdS-to-dS (anti-de Sitter to de Sitter) transition in the vacuum energy: the effective cosmological constant $\Lambda_{\rm s}$ flipped from negative to positive while preserving its magnitude (``mirror'' reflects this invariance), with all other standard components, including baryons, CDM, pre-recombination physics, and the inflationary paradigm, left unaltered.
The idea was originally conjectured phenomenologically in Ref.~\cite{Akarsu:2019hmw}, motivated by hints from the graduated dark energy (gDE) model, where a swift yet smooth mirror AdS-to-dS-like transition passage around $z\!\sim\!2$ helps organize late-time discrepancies (e.g., the $H_0$ and BAO Ly$\alpha$ discrepancies) without disturbing CMB-era physics. Phenomenologically, such transitions can be described with sigmoid-like profiles; for example, $\Lambda_{\rm s}(z)=\Lambda_{\rm s0}\,\frac{\tanh[\eta\,(z_\dagger - z)]}{\tanh(\eta z_\dagger)}$, where $\eta>1$ controls the sharpness, $\Lambda_{\rm s0}>0$ is the present-day value, and $z_\dagger$ denotes the center of the transition~\cite{Akarsu:2022typ}.
For sufficiently rapid transitions (e.g., $\eta\gtrsim10$ at $z_\dagger\!\sim\!2$), this approaches a smooth step with $\Lambda_{\rm s}\!\approx\!\Lambda_{\rm s0}$ for $z\!\lesssim\!2$ and $\Lambda_{\rm s}\!\approx\!-\Lambda_{\rm s0}$ for $z\!\gtrsim\!2$, effectively leaving a single parameter, $z_\dagger$~\cite{Akarsu:2022typ,Akarsu:2024eoo}. In the instantaneous limit $\eta\!\to\!\infty$, $\Lambda_{\rm s}(z)\;\to\;\Lambda_{\rm s0}\,{\rm sgn}(z_\dagger - z)$, which defines the \textit{abrupt} \LsCDM\ model~\cite{Akarsu:2021fol,Akarsu:2022typ,Akarsu:2023mfb}, an idealized representation of a rapid mirror AdS-to-dS transition that extends \LCDM\ by a single additional parameter, $z_\dagger$.
This simplest phenomenological realization of \LsCDM\ has been extensively studied under the assumption of GR; see, e.g., Refs.~\cite{Akarsu:2021fol,Akarsu:2022typ,Akarsu:2023mfb,Yadav:2024duq,Akarsu:2024eoo,Escamilla:2025imi,Paraskevas:2024ytz,Akarsu:2025ijk}. The abrupt limit serves as a controlled proxy for a fast transition, and by construction the limit $z_\dagger\!\to\!\infty$ exactly recovers \LCDM. We refer the reader (without claiming completeness) to the broader literature on \LsCDM\ and cognate scenarios, including theoretical developments, observational constraints, and related frameworks that invoke a negative CC or DE sectors (effective or field-based) admitting negative energy densities at high redshifts, as well as model-agnostic reconstructions pointing in that direction; see Refs.~\cite{Sahni:2002dx, Vazquez:2012ag, BOSS:2014hwf, Sahni:2014ooa, BOSS:2014hhw, DiValentino:2017rcr, Mortsell:2018mfj, Poulin:2018zxs, Capozziello:2018jya, Wang:2018fng, Banihashemi:2018oxo, Dutta:2018vmq, Banihashemi:2018has, Akarsu:2019ygx, Li:2019yem, Visinelli:2019qqu, Ye:2020btb, Perez:2020cwa, Akarsu:2020yqa, Ruchika:2020avj, DiValentino:2020naf, Calderon:2020hoc, Ye:2020oix, DeFelice:2020cpt, Paliathanasis:2020sfe, Bonilla:2020wbn, Acquaviva:2021jov, Bag:2021cqm, Bernardo:2021cxi, Escamilla:2021uoj, Sen:2021wld, Ozulker:2022slu, DiGennaro:2022ykp, Akarsu:2022lhx, Moshafi:2022mva, Bernardo:2022pyz, vandeVenn:2022gvl, Ong:2022wrs, Tiwari:2023jle, Malekjani:2023ple, Vazquez:2023kyx, Escamilla:2023shf, Adil:2023exv, Alexandre:2023nmh, Adil:2023ara, Paraskevas:2023itu, Gomez-Valent:2023uof, Wen:2023wes, Medel-Esquivel:2023nov, DeFelice:2023bwq, Anchordoqui:2023woo, Menci:2024rbq, Anchordoqui:2024gfa,Akarsu:2024qsi, Gomez-Valent:2024tdb, DESI:2024aqx, Bousis:2024rnb, Wang:2024hwd, Colgain:2024ksa, Tyagi:2024cqp, Toda:2024ncp, Sabogal:2024qxs, Dwivedi:2024okk, Escamilla:2024ahl, Anchordoqui:2024dqc, Akarsu:2024nas, Gomez-Valent:2024ejh, Manoharan:2024thb, Souza:2024qwd, Pai:2024ydi,Paraskevas:2024ytz,Akarsu:2025ijk, Mukherjee:2025myk, Giare:2025pzu, Keeley:2025stf, Akarsu:2025gwi, Soriano:2025gxd, Sabogal:2025mkp, Mukherjee:2025ytj, Efstratiou:2025xou, Escamilla:2025imi, Silva:2025hxw, Specogna:2025guo, Scherer:2025esj, Wang:2025dtk, Bouhmadi-Lopez:2025ggl, Tamayo:2025xci, Gonzalez-Fuentes:2025lei, Bouhmadi-Lopez:2025spo, Hogas:2025ahb, Yadav:2025vpx, Lehnert:2025izp, Tan:2025xas, Pedrotti:2025ccw, Forconi:2025gwo, Nyergesy:2025lyi,Ghafari:2025eql}.

In \LsCDM, because the pre-recombination universe remains unaltered, CMB distance anchoring \emph{forces} a compensating late-time shift that \emph{raises} $H_0$ while \emph{lowering} $\Omega_{\rm m0}$. Consequently, as shown in a series of studies (e.g., ~\cite{Akarsu:2021fol,Akarsu:2022typ,Akarsu:2023mfb,Akarsu:2024eoo,Yadav:2024duq,Escamilla:2025imi}), raising $H_0$ to Supernovae and H0 for the Equation of State of Dark Energy (SH0ES) values naturally suppresses the present-day growth rate $f_0\simeq\Omega_{\rm m0}^{\gamma}$ (while retaining the GR benchmark $\gamma\!\simeq\!0.55$ for the growth index) and the clustering metric $S_8\equiv\sigma_8\sqrt{\Omega_{\rm m0}/0.3}$ (even though $\sigma_8$ increases slightly), thereby enabling \LsCDM\ to simultaneously address the $H_0$, $S_8$, and $\gamma$ tensions \emph{without} invoking modified gravity.
Moreover, the framework remains compatible with BAO Ly$\alpha$ data at $z_{\rm eff}\!\sim\!2.3$, with estimates of the present age of the Universe from the oldest globular clusters, and (with neutrino parameters treated as free~\cite{Yadav:2024duq}) with standard neutrino properties. Notably, a rapid transition near $z_\dagger\!\approx\!1.7$ has been identified as the model’s ``sweet-spot'', where these achievements are realized simultaneously~\cite{Akarsu:2021fol,Akarsu:2022typ,Akarsu:2023mfb,Yadav:2024duq,Akarsu:2024eoo,Escamilla:2025imi}.

Crucially, the departures of the abrupt \LsCDM\ scenario from \LCDM\ are \emph{dynamical} and begin before the transition, namely in the AdS epoch of $\Lambda_{\rm s}$~\cite{Akarsu:2021fol}. They arise in the background expansion and, consequently, in the evolution of perturbations, and hence in structure growth. Specifically, for a transition at $z_\dagger\!\sim\!2$ (as suggested by observational analyses~\cite{Akarsu:2021fol,Akarsu:2022typ,Akarsu:2023mfb,Akarsu:2024eoo}), observable departures are confined primarily to the $z\lesssim3$ universe. By $z\sim3$, the dark energy fraction is only a few percent, $\Omega_{\Lambda}(z\!\sim\!3)\simeq \lvert \Omega_{\Lambda_{\rm s}}(z\!\sim\!3)\rvert \sim \mathcal{O}(10^{-2})$, and rapidly becomes negligible, so that for $3\lesssim z\lesssim z_{\rm eq}\simeq3400$, both models are effectively Einstein–de~Sitter and, for $z>z_{\rm rec}\simeq1100$, reproduce the standard pre-recombination cosmology.
The fundamental difference, relative to \LCDM, is that in the AdS-like CC era ($z>z_\dagger$), the expansion rate $H(z)$ is \emph{lower}, with the deficit growing toward the transition and most pronounced for $z_\dagger<z\!\lesssim\!3$, whereas after the transition (dS-like CC era, $z<z_\dagger$), $H(z)$ is \emph{higher} while closely tracking the \LCDM\ redshift dependence. Taken together, these shifts are expected to have clear and distinct dynamical implications for structure growth.

Accordingly, in the absence of dark energy clustering, subhorizon (pressureless) matter perturbations obey $\ddot\delta+2H\dot\delta-4\pi G\rho_m\,\delta=0$~\cite{Akarsu:2022typ,Akarsu:2025ijk}. Hence, owing to the $H(z)$ evolution described above and \emph{relative} to \LCDM, the Hubble friction term $2H\dot\delta$ is increasingly suppressed across the AdS window (toward the transition), leading to progressively enhanced growth. After the transition, although $\Lambda_{\rm s}$ is dS-like as in \LCDM, $H(z)$ is larger and the friction term becomes more pronounced, thereby damping growth more efficiently throughout that era.
Consequently, while in both models the growth index attains the Einstein–de~Sitter value $\gamma_{\rm EdS}=6/11$ at recombination, in the post-recombination era \LCDM\ has $\gamma(a)>\gamma_{\rm EdS}$, with an excess that increases toward low $z$. In contrast, in \LsCDM\ it remains \emph{below} $\gamma_{\rm EdS}$ throughout the dS-like CC era, with a deficit that strengthens toward $z_\dagger$. At the transition, it jumps sharply and thereafter \emph{raises} smoothly toward low $z$, tracking the \LCDM\ curve at slightly higher values while remaining close to the GR benchmark $\gamma\simeq0.55$ overall (see Ref.~\cite{Akarsu:2025ijk}, Fig.~11).

Correspondingly, relative to \LCDM, \LsCDM\ exhibits a progressively enhanced growth rate $f(a)$ throughout the AdS-like CC era, peaking just before the transition (by $\sim15\%$ for $z_\dagger\!\sim\!2$; see Ref.~\cite{Akarsu:2025ijk}, Fig.~10). At the switch, $f(a)$ drops slightly below the \LCDM\ value; thereafter, in the dS-like CC era, it follows the \LCDM\ redshift dependence while remaining lower.
This behavior of $f(a)$ in \LsCDM\ renders the model a natural candidate to alleviate the recently identified growth-index ($\gamma$) tension in \LCDM~\cite{Nguyen:2023fip}. For context, we briefly outline the mechanism (see Refs.~\cite{Akarsu:2025ijk,Escamilla:2025imi} for details). In \LCDM, adopting the Planck-anchored $\Omega_{\rm m0}=0.32$ together with the GR benchmark $\gamma\simeq0.55$ yields $f_0=\Omega_{\rm m0}^{\gamma}\simeq0.53$ for the present-day universe. By contrast, fitting \gLCDM\ (i.e., \LCDM\ with $\gamma$ free) to \mbox{CMB+$f\sigma_8$} data gives $\gamma=0.639\pm0.025$ and $\Omega_{\rm m0}=0.308\pm0.007$, which together imply $f_0=0.471\pm0.015$~\cite{Nguyen:2023fip,Escamilla:2025imi}.
This $\sim4\sigma$ upward shift in $\gamma$ relative to the GR expectation constitutes the growth-index tension and corresponds to a suppressed present-day growth rate ($f_0\simeq0.47$) relative to the Planck-\LCDM\ expectation ($f_0\sim0.53$). By contrast, \LsCDM, at its sweet-spot $z_\dagger\!\sim\!1.7$ and retaining the GR benchmark $\gamma\simeq0.55$, with Planck-anchored $\Omega_{\rm m0}\simeq0.27$~\cite{Akarsu:2021fol,Akarsu:2022typ,Akarsu:2023mfb,Akarsu:2024eoo}, already predicts a suppressed present-day growth rate $f_0\simeq0.49$, closely matching the $f_0\simeq0.47$ inferred in \gLCDM\ analyses of \mbox{CMB+$f\sigma_8$} data.
Indeed, constraining \gLsCDM\ (i.e., \LsCDM\ with free $\gamma$) with the same \mbox{CMB+$f\sigma_8$} data yields $\gamma=0.586\pm0.022$ and $\Omega_{\rm m0}=0.265\pm0.007$, implying $f_0=0.461\pm0.015$—in excellent agreement (within $\sim0.5\sigma$) with the \gLCDM\ inference for $f_0$—while $\gamma$ remains within $\sim1.5\sigma$ of the GR value, thereby alleviating the growth-index tension~\cite{Escamilla:2025imi}.
In \gLCDM, the suppression of $f_0$ inferred from \mbox{CMB+$f\sigma_8$} is achieved by raising the growth index to $\gamma\!\sim\!0.64$, suggestive of a departure from GR, while keeping $\Omega_{\rm m0}$ close to its Planck-\LCDM\ value. In \gLsCDM\ (with $z_\dagger\!\sim\!1.7$), it arises chiefly from a lower matter density—$\Omega_{\rm m0}$ smaller by $\sim0.04$ relative to Planck-\LCDM—while the fitted $\gamma\!\sim\!0.59$ remains statistically consistent with GR ($\gamma\simeq0.55$).

Finally, halo-scale analyses based on spherical collapse in the \emph{abrupt} \LsCDM\ background show that bound structures survive the transition, and that the \emph{virial overdensity} shifts relative to Planck-\LCDM\ depending on whether the switch occurs before or after turnaround—that is, halos virialize with either increased or reduced overdensity contingent on the transition timing. This behavior encodes both the \emph{duration} and \emph{depth} of the pre-transition AdS phase, together with post-transition dS-like CC damping~\cite{Paraskevas:2024ytz}.
In this idealized limit, the transition manifests as a type II (sudden) singularity~\cite{Barrow:2004xh} at $z=z_\dagger$~\cite{Paraskevas:2024ytz}. Crucially, even in its most extreme, abrupt form, the mirror AdS-to-dS transition neither dissociates bound systems nor significantly perturbs Newtonian virialized halos, and this singularity, which already has mild dynamical impact, disappears once the transition is smoothed~\cite{Paraskevas:2024ytz}.\footnote{For a smooth AdS-to-dS transition modeled with a DE field in GR, the type II singularity is absent. Because $p_{\rm DE}$ remains finite while $\rho_{\rm DE}$ crosses zero continuously at $z=z_\dagger$, the equation-of-state parameter $w_{\rm DE} \equiv p_{\rm DE}/\rho_{\rm DE}$ diverges there, $\lim_{z\to z_\dagger^\pm} w_{\rm DE}(z)=\pm\infty$, which, however, is a safe singularity: all background quantities ($a$, $H$, $\dot H$), as well as the total energy density and pressure, remain finite and continuous. In simple scalar-field realizations, the sound speed is luminal, $c_{\rm s}^2 = 1$. See Refs.~\cite{Ozulker:2022slu,Paraskevas:2024ytz,Akarsu:2025gwi} for details.}

Taken together, the linear-regime growth and halo-scale spherical-collapse results point to a concrete, falsifiable \emph{nonlinear} signature characteristic of the \LsCDM\ scenario. During the AdS-like CC era, reduced Hubble friction (relative to \LCDM) seeds an early excess of power at high comoving wavenumbers $k$ (units $h\,\mathrm{Mpc}^{-1}$). After the AdS-to-dS transition, $H(z)$ in the dS-like CC era closely tracks the \LCDM\ redshift dependence but at higher values, thereby suppressing subsequent growth more efficiently than in \LCDM.
As nonlinearity propagates to progressively larger physical scales (i.e., as the nonlinear scale $k_{\rm NL}(z)$ decreases), the \emph{maximum relative difference}, $P_{\LsCDM}/P_{\LCDM} - 1$, is expected to track $k_{\rm NL}(z)$. This appears as a \emph{localized}, redshift-dependent crest in $P_{\LsCDM}/P_{\LCDM}$ that drifts to lower $k$ with time. Unlike a scale-independent rescaling (e.g., of $\sigma_8$ or $A_s$), the crest’s \emph{amplitude} and \emph{$k$-location} encode the integrated background history—how long and how strongly the Universe evolved in the AdS-like CC era, and how rapidly it was subsequently damped in the dS-like CC era.
The crest is expected to arise near the AdS-to-dS transition, initially around $k_{\rm NL}(z_\dagger)$ with, for example, $z_\dagger\!\sim\!1.7$, and then migrate, in comoving $k$, toward the group or poor-cluster regime by $z=0$, where weak lensing, galaxy–galaxy lensing, cluster counts, and tSZ power are most sensitive. If the transition occurs earlier, for example $z_\dagger\!\sim\!2.0$, the same morphology should be obtained but with reduced amplitude and earlier onset: the dark energy fraction is smaller at higher redshift, and the longer post-transition era provides more time to damp and redistribute the pre-transition excess, yielding a weaker residual feature by $z=0$.

\emph{This work} tests these predictions in the fully nonlinear regime using relativistic $N$-body simulations with \texttt{gevolution}. We simulate the \emph{abrupt} \LsCDM\ limit, a controlled proxy for a smooth but rapid transition, alongside a matched \LCDM\ baseline, adopting two independent best-fit parameter sets (Planck-only and a combined ``full'' dataset; see~\cref{tab:bestfit}). Baryons and CDM are co-evolved as a single particle ensemble in a $(2080\,{\rm Mpc}/h)^3$ volume with $0.5\,{\rm Mpc}/h$ spatial resolution; a single massive neutrino with $\sum m_\nu=0.06\,\mathrm{eV}$ is included in the matter sector and evolved linearly. 
We analyze the absolute matter power spectra up to $k_f\equiv k_{\rm Ny}/4\simeq1.6\,h\,\mathrm{Mpc}^{-1}$, and we track the localized feature in their relative behaviour down to $k\gtrsim2\,h\,\mathrm{Mpc}^{-1}$, exploiting reduced sensitivity of the \emph{ratio} $P_{\Lambda_{\rm s}\mathrm{CDM}}/P_{\Lambda\mathrm{CDM}}$ to resolution limits. Consistency in the linear regime is verified by cross-checking \texttt{gevolution} outputs against the Cosmic Linear Anisotropy Solving System (\texttt{CLASS}) in Newtonian gauge.

The paper is organized as follows. Following a review of the background dynamics and linear phenomenology of $\Lambda_{\rm s}{\rm CDM}$ in the Introduction (\cref{sec:intro}), Section~\ref{sec:setup} introduces the abrupt $\Lambda_{\rm s}{\rm CDM}$ background and linear phenomenology, and details our simulation setup and validation. Section~\ref{sec:results} presents the nonlinear power-spectrum results, quantifying the crest’s amplitude and redshift-dependent $k$-drift, and mapping them to observational targets. Section~\ref{sec:conclusion} summarizes the implications and outlines future tests.

\section{Background and simulation setup}
\label{sec:setup}

To study the matter power spectrum in $\Lambda_{\rm s}$CDM relative to $\Lambda$CDM, we perform two sets of $N$-body simulations using the best-fit cosmological parameters from the \emph{Planck}-only and full datasets listed in~\cref{tab:bestfit}. Simulations are carried out with the particle-mesh (PM) relativistic code \texttt{gevolution}~\cite{Adamek:2015eda,Adamek:2016zes}, which by default evolves perturbations in the Poisson gauge. Our use of \texttt{gevolution} is motivated by robustness and consistency: it provides a controlled GR framework in a well-defined gauge, evolves the metric potentials sourcing particle motion self-consistently, and allows a clean inclusion of relativistic species (here implemented linearly).

Massive neutrinos are included in the minimal-mass normal hierarchy with $\sum m_\nu = 0.06\,\mathrm{eV}$, implemented as a single massive eigenstate. Identical clustering is assumed for baryons and CDM (i.e., no bias), and our results do not incorporate baryonic feedback effects on small scales. Baryons and CDM are co-evolved in a single ensemble of $4160^3$ particles, with initial perturbations generated from a weighted average of their separate transfer functions (the \textit{blend} option in \texttt{gevolution}). All runs are performed in boxes of size $L=2080\,{\rm Mpc}/h$ with $4160$ grid points. Additional parameters not shown in~\cref{tab:bestfit} include $N_{ur}=2.0328$ for two ultra-relativistic neutrinos and $T_{\nu}/T_{\gamma}=0.71611$, which are common to both cosmological models.

Outputs for $\Lambda$CDM and $\Lambda_{\rm s}$CDM cosmologies are directly comparable, since all runs within a given dataset (\emph{Planck}-only vs.\ full) share identical simulation and IC (initial conditions) generation settings. The code employs a Fast Fourier Transform (FFT) algorithm and therefore operates at fixed resolution. However, limitations associated with the fixed lattice are partially mitigated when considering the \emph{ratios} $P_{\Lambda_{\rm s}\mathrm{CDM}}/P_{\Lambda\mathrm{CDM}}$, which remain sufficiently reliable for our purposes beyond the $k$-range adopted for the individual absolute spectra $P_{\Lambda_{\rm s}\mathrm{CDM}}$ and $P_{\Lambda\mathrm{CDM}}$.

Initial perturbation amplitudes at $z_{\rm ini} = 100$ are generated with the Boltzmann code \texttt{CLASS}~\cite{Blas:2011rf}. In the simulations, metric evolution includes massive-neutrino perturbations only at linear order, in contrast to the joint CDM-baryon component (hereafter denoted cb), which is represented by an $N$-body ensemble. Time-dependent neutrino transfer functions are provided by \texttt{CLASS} at runtime~\cite{Adamek:2017uiq,Brandbyge:2008js}.

\begin{table}[t]
\centering
\caption{Best-fit cosmological parameters for the $\Lambda$CDM and $\Lambda_{\rm s}$CDM models used in our simulations. Values are derived from the parameter analyses of Ref.~\cite{Akarsu:2023mfb}: ``Planck'' corresponds to \emph{Planck} CMB-only, while ``Full'' denotes the combined set \emph{Planck}+BAO$_{\rm tr}$+PantheonPlus\&SH0ES+KiDS-1000. The values listed are nearby best-fits obtained by locally re-optimizing around the posteriors of Ref.~\cite{Akarsu:2023mfb} (see~\cref{sec:setup}).}
\label{tab:bestfit}
\scalebox{0.75}{%
\begin{tabular}{lcccc}
\hline\hline
Parameter & $\Lambda$CDM (Planck) &$\Lambda$CDM (full) & $\Lambda_{\rm s}$CDM (Planck) & $\Lambda_{\rm s}$CDM (full) \\
\hline
$10^2\omega_b$              & $2.235$  & $2.283$  & $2.236$ & $2.243$ \\
$\omega_{\rm cdm}$          & $0.1202$ & $0.1143$ & $0.1202$ & $0.1192$ \\
$100\,\theta_*$             & $1.0418$ & $1.0422$ & $1.0417$ & $1.0418$ \\
$\ln(10^{10}A_s)$           & $3.049$  & $3.079$  & $3.036$ & $3.030$ \\
$n_s$                       & $0.9658$ & $0.9797$ & $0.9648$ & $0.9651$ \\
$\tau$                      & $0.0549$ & $0.0771$ & $0.0484$ & $0.0485$ \\
$z_\dagger$                 & ---      & ---      & $1.927$ & $1.669$ \\
\hline
$\Omega_{\rm m}$                  & $0.3163$ & $0.2816$ & $0.2872$ & $0.2676$ \\
$\sigma_8$                  & $0.8136$ & $0.8076$ & $0.8216$ & $0.8238$ \\
$H_0\,[\mathrm{km\,s^{-1}Mpc^{-1}}]$ & $67.28$  & $69.96$  & $70.61$ & $72.90$ \\
$t_0\,[\mathrm{Gyr}]$       & $13.80$  & $13.71$  & $13.62$ & $13.53$ \\
\hline
\end{tabular}}
\end{table}

The codes are left unaltered for \LCDM. For \LsCDM, we modify only the \emph{background} evolution in both \texttt{gevolution} and \texttt{CLASS} to implement a non-clustering, sign-switching cosmological constant $\Lambda_{\rm s}$. In this framework, the standard cosmological constant $\Lambda$ of the \LCDM\ model is replaced by $\Lambda_{\rm s}$, which undergoes a rapid sign switch near $z_\dagger$, with an AdS-like phase ($\Lambda_{\rm s} < 0$) at earlier times and a dS-like phase ($\Lambda_{\rm s} > 0$) thereafter. We adopt the idealized step-limit prescription
\begin{equation}
\Lambda_{\rm s}(z) = \Lambda_{\rm s0} \,\mathrm{sgn}(z_\dagger - z)\,,
\end{equation}
where $\mathrm{sgn}$ denotes the signum function~\cite{Akarsu:2021fol,Akarsu:2022typ,Akarsu:2023mfb}. This choice modifies only the background expansion (i.e. no dark energy clustering) and serves as a simplifying approximation to a fast but continuous sign switch. The Friedmann equation then reads
\begin{equation}
\frac{H^2}{H_0^2}
= \Omega_{\rm r0}(1+z)^4 + \Omega_{\rm m0}(1+z)^3
+ \Omega_{\Lambda_{\rm s}0} \,\mathrm{sgn}(z_\dagger - z)\, .
\end{equation}
The present-day density parameters retain their standard definitions (setting $c \equiv 1$)
${\Omega_{\rm r0} = 8\pi G\,\varepsilon_{\rm r0}/(3H_0^2)}$,
${\Omega_{\rm m0} = 8\pi G\,\varepsilon_{\rm m0}/(3H_0^2)}$, and
${\Omega_{\Lambda_{\rm s}0} = \Lambda_{\rm s0}/(3H_0^2)}$. On subhorizon scales, the pressureless linear growth still satisfies~\cite{Akarsu:2022typ}
\begin{equation} \label{growth}
\ddot\delta + 2H\dot\delta - 4\pi G\rho_m\,\delta = 0 \, .
\end{equation}
For the baseline free parameters we rely on the recent global analysis of \LsCDM\ in Ref.~\cite{Akarsu:2023mfb}, which identifies a transition at $z_\dagger \simeq 1.7$–$1.9$ and reports improvements across multiple data combinations relative to \LCDM.

It has been shown in Ref.~\cite{Adamek:2017grt} that in \texttt{gevolution} simulations carried out in the Poisson gauge, the impact of early radiation on the matter power spectrum—specifically on very large scales—remains at the sub-percent level. Motivated by this, we include photon and massless-neutrino perturbations only down to $z = 15$. Their transfer functions are computed with \texttt{CLASS} and supplied to \texttt{gevolution} at runtime to construct the corresponding density fields (analogous to the treatment of massive neutrinos). At later times, down to $z = 0$, these species are retained only in the background evolution.

The entries in~\cref{tab:bestfit} are point estimates obtained by locally re-optimizing the likelihood around the MontePython chains of Ref.~\cite{Akarsu:2023mfb} for the two datasets employed here: ``Planck'' (Planck CMB only) and ``full'' (the combined \emph{Planck}+BAO$_{\rm tr}$+PantheonPlus\&SH0ES+KiDS-1000 analysis). Ref.~\cite{Akarsu:2023mfb} sampled parameters with \texttt{CLASS}+MontePython (Metropolis-Hastings, $R - 1 < 10^{-2}$) and reported 68\%\,CL posteriors and best fits; our local re-optimization yields nearby best fits consistent with those posteriors. Small numerical shifts relative to Ref.~\cite{Akarsu:2023mfb} are immaterial and leave all figures unchanged at the precision shown.

Throughout, ``total matter'' denotes the sum of baryons, CDM, and massive neutrinos. Comparisons to linear theory are performed with \texttt{CLASS} in Newtonian gauge (the gauge choice is discussed below).

\section{Results and interpretation}
\label{sec:results}

\begin{figure}[t!]
    \centering
    \includegraphics[width=1.0\linewidth]
    {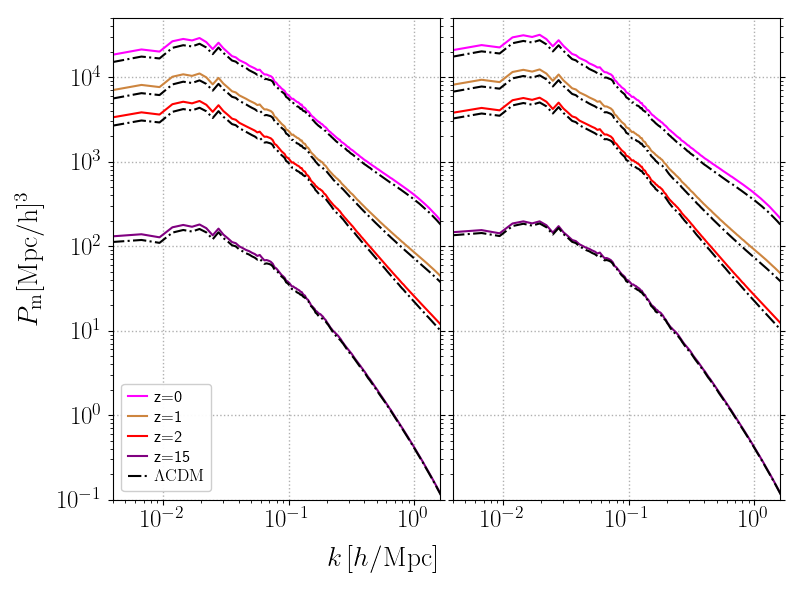}
    \caption{Total-matter (baryons+CDM+massive $\nu$) power spectra from our \texttt{gevolution} simulations, shown for the \emph{Planck}-only (left) and full (right) best-fit parameter sets at $z=15,2,1,0$. Colored solid curves correspond to \LsCDM, while black dash-dotted curves indicate \LCDM. Within each dataset, both runs share identical simulation and IC generation settings. Absolute spectra are plotted up to $k_{\rm f} = 1.6\,h\,\mathrm{Mpc}^{-1} \simeq k_{\rm Ny}/4$.}
    \label{planck2}
\end{figure}

\begin{figure}[t!]
    \centering
    \includegraphics[width=1.0\linewidth]{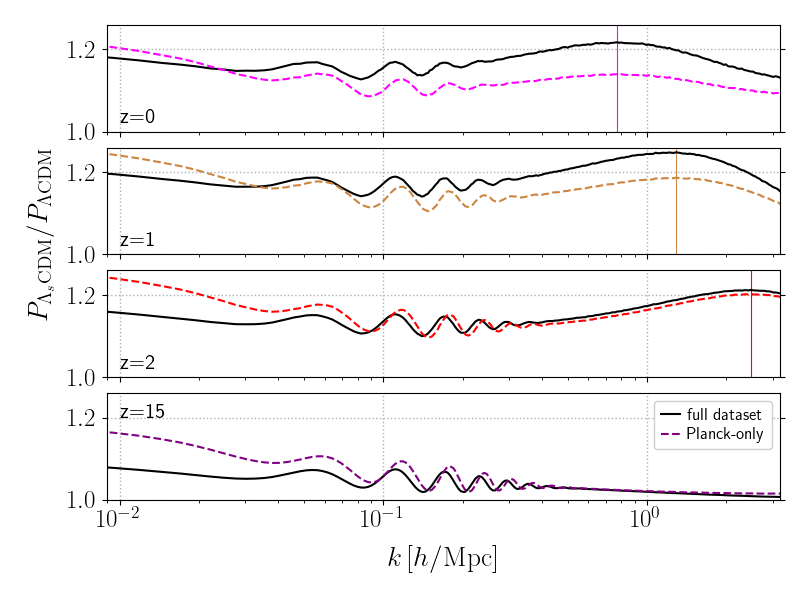}
    \caption{Ratios of the total-matter power spectra, $P_{\Lambda_{\rm s}\mathrm{CDM}}/P_{\Lambda\mathrm{CDM}}$, at $z=15,2,1,0$ for the \emph{Planck}-only (dashed) and full (solid) best-fit parameter sets. Vertical tick marks denote the $k$-locations of the local maxima from $z=2$ onward. Peak amplitudes are: \emph{Planck}-only, $1.202$ at $z=2$, $1.186$ at $z=1$, $1.140$ at $z=0$; full dataset, $1.212$ at $z=2$, $1.248$ at $z=1$, $1.217$ at $z=0$.}
        \label{Pl-fd}
\end{figure}

\cref{planck2} presents the total-matter power spectra at $z=15,2,1,0$ for both cosmologies, using the \emph{Planck}-only (left) and full (right) best-fit parameter sets in~\cref{tab:bestfit}. Approaching $k \sim \mathcal{O}(1)\,h\,\mathrm{Mpc}^{-1}$, at very high redshift ($z=15$), the two models are nearly indistinguishable, as expected when the the dark energy fraction is negligible ($\Omega_{\Lambda}\simeq \lvert \Omega_{\Lambda_{\rm s}}\rvert\approx0$); the small residual offset in power reflects the different late-time best fits (e.g., $A_s$, $n_s$, $\Omega_{\rm m0}$). At lower redshift, in the nonlinear regime of interest ($k \gtrsim 0.1\,h\,\mathrm{Mpc}^{-1}$), the largest \emph{relative} difference appears in the neighborhood of the sign-switch (\textit{mirror} AdS-to-dS transition) epoch and follows the dataset-dependent $z_\dagger$: for the \emph{Planck}-only case it is maximal at $z \simeq 2$ (just before $z_\dagger \simeq 1.93$), whereas for the full dataset it is maximal at $z \simeq 1$ (consistent with $z_\dagger \simeq 1.67$). This behavior is made explicit in~\cref{Pl-fd}, which shows the ratios $P_{\Lambda_{\rm s}\mathrm{CDM}}/P_{\Lambda\mathrm{CDM}}$ at the same four redshifts. The enhancement peaks at wavenumbers $k \simeq 1$--$3\,h\,\mathrm{Mpc}^{-1}$ with amplitudes of order $20$--$25\%$ (precisely: $1.202$ at $z=2$ for \emph{Planck}-only; $1.248$ at $z=1$ for the full dataset), i.e., within the wavelength interval $\lambda = 2\pi/k \simeq 2.1$--$6.3\,h^{-1}\mathrm{Mpc}$ where nonlinear collapse is most efficient. After the sign switch, i.e., in the dS-like CC era ($z < z_\dagger$), the relative excess diminishes but does not vanish: by $z=0$, the crest has migrated to larger physical scales, with its peak at $k \simeq 0.8\,h\,\mathrm{Mpc}^{-1}$ and amplitudes of $1.140$ (\emph{Planck}-only) and $1.217$ (full). In other words, the peak amplitude declines from $\sim\!20\%$ at $z=2$ to $\sim\!14\%$ at $z=0$ for the \emph{Planck}-only case, and from $\sim\!25\%$ at $z=1$ to $\sim\!22\%$ at $z=0$ for the full dataset. The vertical tick marks in~\cref{Pl-fd} indicate the $k$-locations of the local maxima at each redshift; these are essentially insensitive to small parameter differences \emph{within} each dataset, confirming that the peak’s position is controlled by the dynamical history (i.e., the timing of the mirror AdS-to-dS transition) rather than by mild shifts in $(\Omega_{m0},\,H_0,\,A_s,\,n_s)$.


The absolute spectra in~\cref{planck2} are shown only up to $k_{\rm f}=1.6\,h\,\mathrm{Mpc}^{-1}$, defined as one quarter of the Nyquist frequency, $k_{\rm Ny}=\pi/\mathrm{(resolution)}\simeq\pi/0.5\simeq6.3\,h\,\mathrm{Mpc}^{-1}$. The leading-order discretization error relative to the continuum prediction, $P_{\rm continuum}$, partly cancels in the \emph{ratios} $P_{\Lambda_{\rm s}\mathrm{CDM}}/P_{\Lambda\mathrm{CDM}}$, as discussed around Eq.~(5.1) of Ref.~\cite{Adamek:2017uiq}, making them less sensitive to the limitations of fixed resolution. Accordingly, in~\cref{Pl-fd} we report the locations and amplitudes of the maxima at $z=2$ even though they lie slightly beyond $k_{\rm f}$. These should be regarded as robust \emph{relative} features, while absolute amplitudes beyond $k_{\rm f}$ are not interpreted further here.

Figures~\ref{class-lscdm} (\emph{Planck}-only) and \ref{class} (full dataset) compare the cb-only power spectra from $N$-body runs (solid) to the corresponding linear predictions from \texttt{CLASS} (black dash-dotted). As noted earlier, although the cb component shown separately here \emph{excludes} massive $\nu$, their perturbations still source the metric linearly. On large, linear scales ($k\!\lesssim\!0.1$–$0.2\,h\,\mathrm{Mpc}^{-1}$) the agreement between \texttt{gevolution} and \texttt{CLASS} is excellent at all redshifts, validating the consistent implementation of the modified background in the \LsCDM\ cosmology across both codes. We emphasize that in \texttt{CLASS} we work in Newtonian gauge, since calculations in the default synchronous gauge are sensitive to the step in $H(z)$ for the abrupt \LsCDM\ model. Meanwhile, our interpretation of the linear-theory versus simulation comparison focuses on subhorizon modes, $k \gtrsim 0.02\,h\,\mathrm{Mpc}^{-1}$, where finite-volume effects are negligible and gauge effects are no longer significant. For a complementary benchmark against an analytic nonlinear prescription (\texttt{HMCode}) and an estimate of its range of validity in this setting, see Appendix~\ref{app:hmcode} (Figs.~\ref{model_ratios} and~\ref{method_ratios}). Toward smaller scales the $N$-body spectra exceed the linear curves at low redshift, as expected from nonlinear growth. The corresponding \emph{ratios} in Figs.~\ref{planckvs-cl} and \ref{fdcl} show the same trend: linear-theory ratios remain smooth and modest, whereas their $N$-body counterparts develop a localized crest around $k\!\sim\!\mathcal{O}(1)\,h\,\mathrm{Mpc}^{-1}$ at $z\!=\!2,1,0$ (earlier and weaker for the \emph{Planck}-only fit, later and stronger for the full dataset), confirming that the enhancement is a genuine nonlinear response to the \LsCDM\ background.

A scale-independent change to the linear normalization (e.g., tuning $\sigma_8$ or $A_s$) cannot reproduce the \emph{localized}, redshift-dependent crest that we find in the simulation ratio $P_{\LsCDM}/P_{\LCDM}$ (Figs.~\ref{Pl-fd},~\ref{planckvs-cl},~\ref{fdcl}). In our results the excess is confined to $k\sim\mathcal{O}(1)\,h\,{\rm Mpc}^{-1}$ and its peak shifts to larger physical scales (i.e., smaller comoving $k$) toward $z=0$, reaching $k\simeq0.6$–$1.0\,h\,{\rm Mpc}^{-1}$, while large scales remain consistent with linear-theory predictions in Newtonian (Poisson) gauge (cf.\ Figs.~\ref{class-lscdm}-\ref{class}). A uniform amplitude change would affect all $k$ similarly and cannot generate a drifting, localized crest; the feature is therefore \emph{dynamical}, tracing the sign-switch (the mirror AdS-to-dS transition in $\Lambda_{\rm s}$) history of $H(z)$ rather than a static renormalization of the spectrum.

\subsection*{Physical origin of the \texorpdfstring{\LsCDM}{\(\Lambda_{\rm s}\)CDM}–\texorpdfstring{\LCDM}{\(\Lambda\)CDM} deviations across redshift}

\begin{figure}[t!]
    \centering
    \includegraphics[width=1.0\linewidth]
    {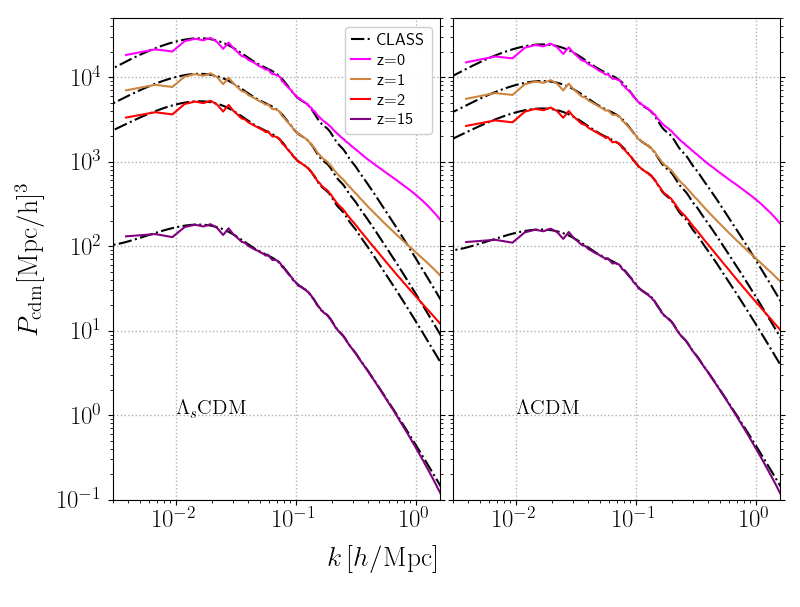}
    \caption{CDM-baryon (cb) power spectra from \texttt{gevolution} at $z=15,2,1,0$ for the \emph{Planck}-only best-fit parameters. Dash-dotted curves show the corresponding \emph{linear} predictions from \texttt{CLASS} in Newtonian gauge. Agreement is excellent on large scales, while deviations at higher $k$ reflect nonlinear clustering for $z \leq 2$.}
    \label{class-lscdm}
\end{figure}

\begin{figure}[t!]
    \centering
    \includegraphics[width=1.0\linewidth]
    {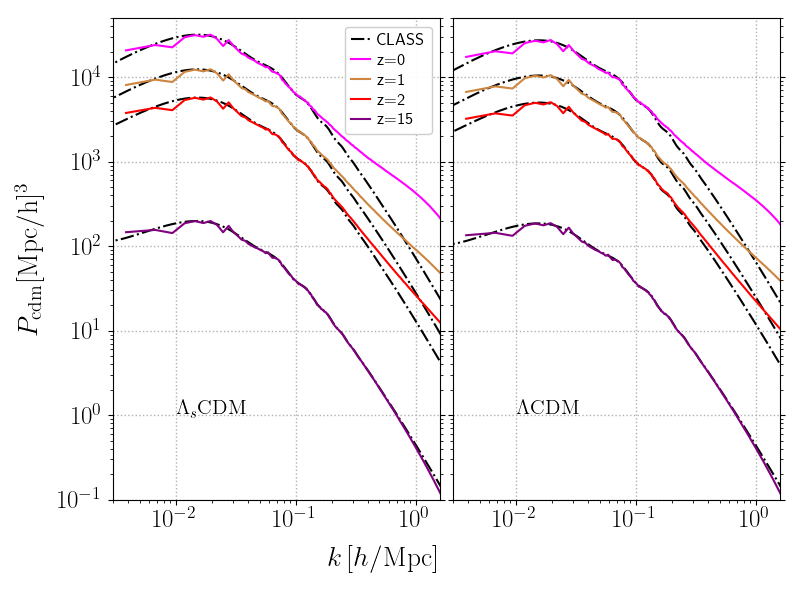}
    \caption{CDM-baryon (cb) power spectra from \texttt{gevolution} at $z=15,2,1,0$ for the full best-fit parameters. Dash-dotted curves show the corresponding \emph{linear} predictions from \texttt{CLASS} in Newtonian gauge. As in the \emph{Planck}-only case, excellent agreement is found on large scales between each $N$-body and linear prediction pair.}
    \label{class}
\end{figure}

\begin{figure}[t!]
    \centering
    \includegraphics[width=1.0\linewidth]{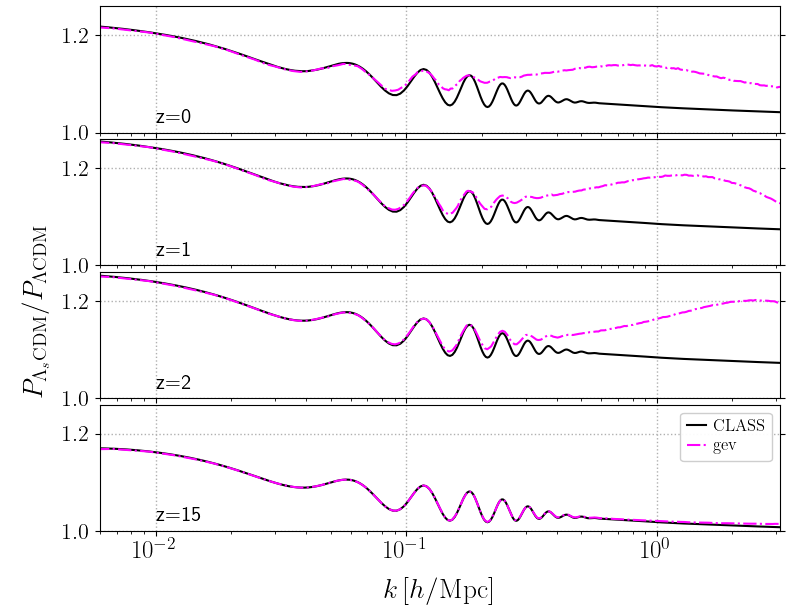}
        \caption{Ratios of CDM--baryon (cb) power spectra, $P_{\Lambda_{\rm s}{\rm CDM}}/P_{\Lambda{\rm CDM}}$, at $z=15,2,1,0$ for the \emph{Planck}-only best-fit parameters. A localized enhancement appears near the switch epoch and subsequently drifts to larger physical scales, reaching $k\!\sim\!0.6$–$1.0\,h\,\mathrm{Mpc}^{-1}$ by $z=0$. On linear scales the relative spectra from simulations agree with those from \texttt{CLASS}, as expected.}
        \label{planckvs-cl}
\end{figure}

\begin{figure}[t!]
    \centering
    \includegraphics[width=1.0\linewidth]{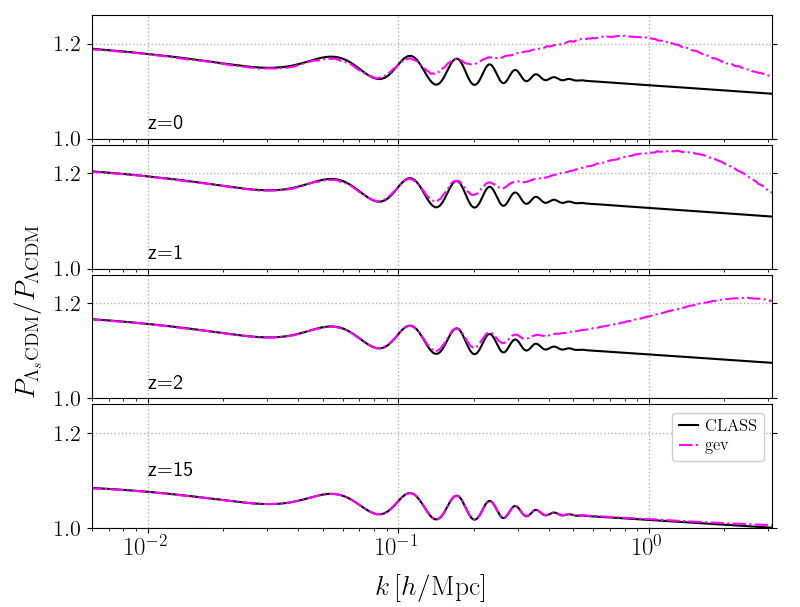}
       \caption{Ratios of CDM-baryon (cb) power spectra, $P_{\Lambda_{\rm s}{\rm CDM}}/P_{\Lambda{\rm CDM}}$, at $z=15,2,1,0$ for the full best-fit parameters. As in the \emph{Planck}-only case, a localized enhancement emerges near the switch epoch and drifts to larger physical scales by $z=0$ ($k\!\sim\!0.6$–$1.0\,h\,\mathrm{Mpc}^{-1}$). In the linear regime the simulation ratios agree with those from \texttt{CLASS}.}
        \label{fdcl}
\end{figure}

The pattern we observe follows directly from how the presence of a rapid mirror AdS-to-dS transition in the late universe ($z_\dagger\sim2$) reshapes the Hubble friction term in the (subhorizon, pressureless) linear growth equation given~\cref{growth}, together with subsequent nonlinear mode coupling. At very high redshift ($z\gtrsim3.5$),  the dark energy fraction is negligible $\Omega_{\Lambda}(z\!\gtrsim\!3)\simeq \lvert \Omega_{\Lambda_{\rm s}}(z\!\gtrsim\!3)\rvert \lesssim \mathcal{O}(10^{-2})$, so \LsCDM\ and \LCDM\ are practically indistinguishable near $k\sim\mathcal{O}(1)\,h\,\mathrm{Mpc}^{-1}$. The small offsets seen at $z=15$ are indirect, tracing parameter reoptimization (e.g., $A_s$, $n_s$, $\Omega_{\rm m0}$) that fix the growth normalization and transfer-function shape at the initial redshift (cf.\ Figs.~\ref{planck2}, \ref{class-lscdm}, \ref{class}). Approaching the transition from above ($z_\dagger<z\lesssim3.5$), the effective CC is AdS-like ($\Lambda_{\rm s}<0$), the expansion rate is lowered at fixed $(\Omega_m,\Omega_r)$, the friction piece $2H\dot\delta$ is reduced, and the linear growth factor is temporarily enhanced during this window ($D_{\Ls}/D_{\Lambda}>1$). In the simulations, this manifests as a localized, scale-dependent boost peaking near the mirror AdS-to-dS transition, consistent with the dataset-dependent $z_\dagger$ (see the total-matter and cb ratios in Figs.~\ref{Pl-fd}, \ref{planckvs-cl}, \ref{fdcl}).


After the transition ($z<z_\dagger$), the effective CC is dS-like ($\Lambda_{\rm s}>0$) as in \LCDM, but the best-fit \LsCDM\ backgrounds (both the \emph{Planck}-only and full) feature a larger late-time expansion rate, which increases Hubble friction and partially suppresses subsequent growth. Crucially, this acts on a field that was already amplified during the AdS-like effective CC phase ($\Lambda_{\rm s}<0$), so the imprint cannot be erased. Nonlinear evolution then transfers part of the excess toward larger physical scales: in Fourier space, the crest in $P_{\Ls}/P_{\Lambda}$ migrates from $k\simeq2$--$3\,h\,\mathrm{Mpc}^{-1}$ ($k\simeq1$--$2\,h\,\mathrm{Mpc}^{-1}$ for the full dataset) near the transition to $k\simeq0.6$--$1.0\,h\,\mathrm{Mpc}^{-1}$ by $z=0$,\footnote{This interval roughly brackets a flat top at $z=0$ in~\cref{Pl-fd} for both datasets, with the local maxima near $k\simeq0.8\,h\,\mathrm{Mpc}^{-1}$ in both total-matter and cb ratios (Figs.~\ref{Pl-fd}, \ref{planckvs-cl}, \ref{fdcl}).} consistent with mode coupling and the evolving one-halo--two-halo balance. 
The comoving length that corresponds to the top of this crest, $\ell=2\pi/k\simeq 8\,h^{-1}\mathrm{Mpc}$, after mean-density top-hat mapping (whose response peaks near $kR\simeq\pi$), translates to $R\simeq\pi/k\simeq4\,h^{-1}\mathrm{Mpc}$ and yields a characteristic mass $M(R)\equiv(4\pi/3)\,\bar\rho_m\,R^3\sim\mathrm{few}\times10^{13}\,h^{-1}M_\odot$ with $\bar\rho_m=\Omega_{\rm m0}\rho_{\rm c0}$. Across the crest, i.e., $k\simeq0.6$--$1.0\,h\,\mathrm{Mpc}^{-1}$, we get $R\simeq3.1$--$5.2\,h^{-1}\mathrm{Mpc}$ and $M\simeq(1.5$--$6.8)\times10^{13}\,M_\odot$, $M\simeq(1.3$--$6.1)\times10^{13}\,M_\odot$ for the \emph{Planck}-only and full dataset, respectively, i.e., group or poor-cluster scales. 
We emphasize that $M(R)$ here is a \emph{characteristic linear scale} defined from the \emph{background} matter density; virial halo masses at these radii are larger because collapsed regions have overdensities $\Delta\gg1$ (tens to hundreds), and their virial radii ($R_{\rm vir}\sim R\,\Delta^{-1/3}$) are correspondingly smaller than $R$. We do not quote a more precise $M$ because the mapping depends on the chosen window and on halo definitions; the key point is that the affected modes lie in the group or poor-cluster regime, precisely the scales most relevant for galaxy-galaxy lensing, small-scale shear, cluster counts and tSZ statistics. For reference, $\sigma_8$ is the root-mean-square (rms) of the \emph{linear} matter field today smoothed with a top-hat of radius $8\,h^{-1}\mathrm{Mpc}$, whereas the crest we discuss lies at the smaller scale $R\simeq4\,h^{-1}\mathrm{Mpc}$.

Overall, the figures show a coherent causal chain: parameter-driven, percent-level offsets only where the effective CC sector is irrelevant; a genuine dynamical enhancement tied to the AdS-like effective CC window; and a persistent, drifting crest thereafter, viz., in the dS-like CC era, seen clearly in both the total-matter and cb-only spectra, which are consistent with linear theory on large scales and underscore the robustness of this interpretation.

\subsection*{Connection to cosmic noon and observational tests}


Observational reconstructions place the cosmic star-formation-rate density (SFRD) peak at $z\simeq1.5$--$2$ (see, e.g.,~\cite{Madau:2014bja}). In standard $\Lambda$CDM, its timing and amplitude emerge from the interplay of halo assembly, gas accretion, cooling and metal enrichment, and feedback; gravity sets the large-scale preconditions but does not by itself fix the peak. In this context, our simulations show that $\Lambda_{\rm s}$CDM supplies a stronger gravitational backdrop near the relevant epoch for \emph{both} data combinations. The maximum near $z\sim2$ for the \emph{Planck}-only fit, for instance, appears at $k\simeq2$--$3\,h\,\mathrm{Mpc}^{-1}$. As worked out in detail in the previous subsection, this interval maps to $R\simeq1.0$--$1.6\,h^{-1}\mathrm{Mpc}$ and hence to $M(R)\sim10^{11}$--$10^{12}\,M_\odot$. For \emph{collapsed} halos, the corresponding virial masses are pushed to $M_{\rm vir}\sim10^{13}$--$10^{14}\,M_\odot$, i.e., the range widely believed to dominate the SFRD.

The localized boost at $k\sim1$--$3\,h\,\mathrm{Mpc}^{-1}$ near $z\sim1$--$2$ aligns with the broader ``cosmic noon'' picture: these modes correspond to the group and cluster halo regime where rapid \emph{halo growth}, \emph{compaction and quenching}, and \emph{black-hole accretion} are observed to peak, with the SFRD maximized at similar epochs (e.g.,~\cite{Madau:2014bja,Barro:2012ru,vanderWel:2014wba,Aird:2015aye,Ueda:2014tma}). The same $k$ range is also where \emph{thermal Sunyaev--Zel'dovich (tSZ)} and \emph{galaxy--galaxy lensing} and \emph{cosmic shear} measurements have their highest sensitivity~\cite{Kilbinger:2014cea,Mandelbaum:2017jpr,Hill:2013dxa,George:2014oba}. We emphasize that cosmic noon itself is not a tension for $\Lambda$CDM once baryonic physics is modeled; rather, our result, in $\Lambda_{\rm s}$CDM, provides a \emph{gravitational prior} that can modestly modulate collapse efficiencies near that epoch. While detailed outcomes (quenching pathways, size growth, AGN duty cycles) are baryon-controlled and beyond the scope of our dark-matter-only simulations, the predicted redshift-dependent and \emph{localized} power excess offers a concrete near-term target for hydrodynamical simulations and for small-scale lensing and tSZ analyses.

\section{Conclusion}\label{sec:conclusion}

We have presented the results of relativistic $N$-body simulations showing how replacing the standard cosmological constant $\Lambda$ of $\Lambda$CDM with a sign-switching cosmological constant~\cite{Akarsu:2019hmw,Akarsu:2021fol,Akarsu:2022typ,Akarsu:2023mfb}, $\Lambda_{\rm s}$---which undergoes a rapid mirror AdS-to-dS transition in the late universe (at redshift $z_\dagger\sim2$)---modifies structure growth in the nonlinear regime, using \texttt{gevolution} with an \textit{abrupt} $\Lambda_{\rm s}$CDM~\cite{Akarsu:2021fol,Akarsu:2022typ,Akarsu:2023mfb} background under general relativity and two independent best-fit parameter sets (\emph{Planck} CMB-only and the ``full'' combination). While $\Lambda_{\rm s}$CDM and $\Lambda$CDM are observationally indistinguishable at $z\gtrsim3$ on $k\sim\mathcal{O}(1)\,h\,\mathrm{Mpc}^{-1}$, the AdS-like CC ($\Lambda_{\rm s}<0$) phase for $z_\dagger<z\lesssim3$ reduces Hubble friction and \emph{dynamically} amplifies the growth of matter perturbations. After the sign switch ($z<z_\dagger$), the CC is dS-like ($\Lambda_{\rm s}>0$) and the late-time expansion rate is larger than in $\Lambda$CDM, partially suppressing subsequent growth but not erasing the pre-amplified field. In Fourier space, this causal sequence yields a localized, redshift-dependent excess: a crest in $P_{\Lambda_{\rm s}{\rm CDM}}/P_{\Lambda{\rm CDM}}$ that peaks near the transition at $k\simeq1$--$3\,\kunit$ with a fit-dependent amplitude of $\simeq1.20$--$1.25$, advected to larger physical scales with time, leaving a robust $15$--$20\%$ uplift at $z=0$ around $k\simeq0.6$--$1.0\,\kunit$ (Figs.~\ref{planck2}, \ref{Pl-fd}). In both absolute and relative cb-only spectra, large-scale results agree with linear predictions (Figs.~\ref{class-lscdm}--\ref{fdcl}), where \texttt{CLASS} comparisons are performed in Newtonian gauge.

The affected wavenumbers correspond to physically distinct, observationally accessible regimes. For the \emph{Planck}-only data (full dataset), near the transition epoch $z\sim2$ ($z\sim1$), the interval $k\simeq 2$-$3\,\kunit$ ($k\simeq 1$-$2\,\kunit$) maps to $R\simeq \pi/k\simeq 1.0$-$1.6\,h^{-1}\mathrm{Mpc}$ ($R\simeq 1.6$-$3.1\,h^{-1}\mathrm{Mpc}$) and a \emph{characteristic Lagrangian} mass $M(R)\simeq 5.4\times10^{11}$-$1.8\times10^{12}\,M_\odot$ ($M(R)\simeq 1.6\times10^{12}$-$1.3\times10^{13}\,M_\odot$) from the mean-density top-hat relation. Interpreted in terms of \emph{collapsed} halos (e.g., $M_{200\mathrm{c}}$ or $M_{500\mathrm{c}}$), the same modes correspond to group-cluster systems with $M\sim10^{13}$-$10^{14}\,M_\odot$ ($M\sim10^{14}$-$10^{15}\,M_\odot$) once typical virial overdensities $\Delta\sim 100$-$200$ are included; this is precisely the mass range widely believed to dominate the SFRD at ``cosmic noon.'' By $z=0$, the migrating crest sits at $k\simeq 0.6$-$1.0\,\kunit$ (comoving scales $\ell=2\pi/k\simeq 6$-$10\,h^{-1}\mathrm{Mpc}$), which corresponds to $R\simeq 3.1$-$5.2\,h^{-1}\mathrm{Mpc}$ and $M(R)\simeq (1.5$-$6.8)\times10^{13}\,M_\odot$, $M(R)\simeq (1.3$-$6.1)\times10^{13}\,M_\odot$ for the \emph{Planck}-only and full dataset, respectively, i.e., group or poor-cluster environments. Unlike a scale-independent shift in the fluctuation summaries $\sigma_8$ or $S_8$ (which largely act as global rescalings), \LsCDM\ predicts a \emph{shape} feature: a localized, redshift-dependent crest whose $k$-location tracks $z_\dagger$ and drifts to larger physical scales with time. Within a given redshift slice, the \emph{Planck}-only and full-dataset runs place the local maxima at nearly the same $k$ (Fig.~\ref{Pl-fd}); their differences lie primarily in the crest’s \emph{amplitude} and in which slice shows the strongest signal (earlier for \emph{Planck}-only, later for full), consistent with their different $z_\dagger$. This signature, if present, should manifest as a localized excess in small-scale cosmic shear and galaxy-galaxy lensing, a tilt in the tSZ power around group scales, and modest, redshift-dependent changes in the abundance and internal structure of halos in the $10^{13}$-$10^{15}\,M_\odot$ range.

Our analysis excludes baryonic physics and scenarios beyond $\sum m_\nu=0.06\,\mathrm{eV}$ in the form of a single massive neutrino. Small-scale amplitudes of the absolute power spectra should be interpreted with caution due to the fixed simulation resolution; accordingly, we show them only up to $k_f\equiv k_{\rm Ny}/4$. Their \emph{ratios}, however, remain informative somewhat beyond $k_f$, enabling us to track the emergence of localized features down to $k\gtrsim 2\,\kunit$. Linear-regime cross-checks further strengthen the robustness of our findings, in particular the consistent implementation of the \LsCDM\ background in the $N$-body framework.

Looking ahead, hydrodynamical simulations and a systematic exploration of $\sum m_\nu$ are required to propagate the crest’s amplitude and $k$-drift into fully marginalized constraints on $(z_\dagger,\Omega_{\Lambda_{\rm s}0})$. On the observational side, a combined analysis of small-scale weak lensing, galaxy-galaxy lensing, cluster counts, and tSZ power offers a sharp test: detecting, or ruling out, a redshift-evolving, localized excess in power at $k\sim\mathcal{O}(1)\,\kunit$ would provide a direct probe of whether late-time cosmic acceleration prefers a sign-switching cosmological constant. In this sense, \LsCDM~\cite{Akarsu:2019hmw,Akarsu:2021fol,Akarsu:2022typ,Akarsu:2023mfb} provides a well-defined benchmark for forthcoming precision probes such as \emph{Euclid}, LSST, and CMB-S4, where small-scale lensing and cluster observables will be key to confirming or excluding a sign-switching vacuum sector.

Another natural direction is to extend our $N$-body program beyond the abrupt \LsCDM~\cite{Akarsu:2021fol,Akarsu:2022typ,Akarsu:2023mfb} approximation adopted under GR. While the abrupt limit provides a clean proof of concept, a fully consistent realization of the scenario calls for a \emph{smooth} transition, replacing the instantaneous sign switch with a well-defined dynamical process. Phenomenologically, this can be modeled with sigmoid-like histories for $\Lambda_{\rm s}(z)$ (e.g., hyperbolic tangents) in redshift (or scale factor), controlled by a rapidity parameter $\eta$ that governs the duration and sharpness of the mirror AdS-to-dS transition (see, e.g.,~\cite{Bouhmadi-Lopez:2025ggl,Bouhmadi-Lopez:2025spo}). One can simulate such smooth families (characterized by $(z_\dagger,\eta)$) within GR and map this two-parameter space onto the crest’s amplitude and $k$-drift in $P_{\Lambda_{\rm s}\mathrm{CDM}}/P_{\Lambda\mathrm{CDM}}$, thereby testing how the \emph{duration}/\emph{rapidity} and functional form of the transition modulate the localized nonlinear signature.
Beyond this phenomenological step, a more fundamental route is to realize the transition via explicit microphysics within GR. A concrete example is Ph-\LsCDM, where a phantom field with a hyperbolic $\tanh$-type potential generates smooth mirror AdS-to-dS transitions and, more generally, asymmetric ones; in this GR framework one expects the growth index to remain near the GR benchmark, $\gamma\simeq0.55$~\cite{Akarsu:2025gwi,Akarsu:2025dmj}. One can also investigate modified gravity embeddings of \LsCDM, in which transient deviations of $\gamma$ can arise around the transition epoch. Representative cases include $\Lambda_{\rm s}$VCDM (a type-II minimally modified gravity)~\cite{Akarsu:2024qsi,Akarsu:2024eoo} and $f(T)$-\LsCDM\ in teleparallel gravity~\cite{Souza:2024qwd} (see also Ref.~\cite{Akarsu:2024nas}). A systematic comparison across these realizations would clarify whether the same small-scale crest favored by data also selects a finite transition duration/rapidity and, if present, a controlled departure from GR confined to the transition epoch.
Finally, there exist theoretical constructions of the \emph{abrupt} scenario itself. $\Lambda_{\rm s}$CDM$^{+}$ provides a string-inspired realization that predicts a modest excess in the total effective number of relativistic species, $N_{\rm eff}\simeq 3.294$~\cite{Anchordoqui:2023woo,Anchordoqui:2024gfa,Anchordoqui:2024dqc}, offering an orthogonal handle via early-time observables (CMB damping tail, BBN). It would also be interesting to explore $\Lambda_{\rm s}$CDM$_{\pm}$~\cite{Soriano:2025gxd}, which extends $\Lambda_{\rm s}$CDM$^{+}$ by allowing the AdS-like CC to have variable depth (as also natural in the Ph--\LsCDM\ framework~\cite{Akarsu:2025gwi}). Such generalizations introduce a degeneracy between the transition redshift and the AdS depth, but they predict richer dynamics in the nonlinear growth regime; in particular, sufficiently large negative AdS values could impact earlier epochs of cosmic history, potentially even before recombination. Both avenues furnish complementary tests: any detection (or null result) of a redshift-evolving, localized crest in the late-time matter power spectrum can be cross-checked against these early-time signatures, tightening the overall assessment of the \LsCDM\ framework.

\acknowledgements

The authors are grateful to Julian Adamek for insightful comments and valuable suggestions. They also thank Emre \"{O}z\"{u}lker for fruitful discussions during the initial stages of this work. Finally, they thank Merab Gogberashvili for drawing their attention to the redshift of the peak in the cosmic star-formation rate.
\"{O}.A. acknowledges the support by the Turkish Academy of Sciences in scheme of the Outstanding Young Scientist Award  (T\"{U}BA-GEB\.{I}P). E.D.V. is supported by a Royal Society Dorothy Hodgkin Research Fellowship. This work was partially supported by the Center for Advanced Systems Understanding (CASUS), financed by Germany's Federal Ministry of Education and Research (BMBF) and the Saxon state government out of the State budget approved by the Saxon State Parliament. Computing resources used in this work were provided by the high-performance computer at the NHR Center of TU Dresden, jointly supported by the Federal Ministry of Education and Research and the state governments participating in the NHR (www.nhr-verein.de/unsere-partner). This article is based upon work from COST Action CA21136 Addressing observational tensions in cosmology with systematics and fundamental physics (CosmoVerse) supported by COST (European Cooperation in Science and Technology). 

\section*{Data Availability}
Datasets generated to produce the matter power spectra analyzed in this work are publicly available in the Rossendorf Data Repository (RODARE) and may be accessed via \href{https://doi.org/10.14278/rodare.4019}{doi.org/10.14278/rodare.4019} \cite{data}.

\appendix
\section{Benchmark against nonlinear prescriptions: \texttt{HMCode}}
\label{app:hmcode}

\begin{figure*}[t!]
  \centering

  \includegraphics[width=0.495\textwidth]{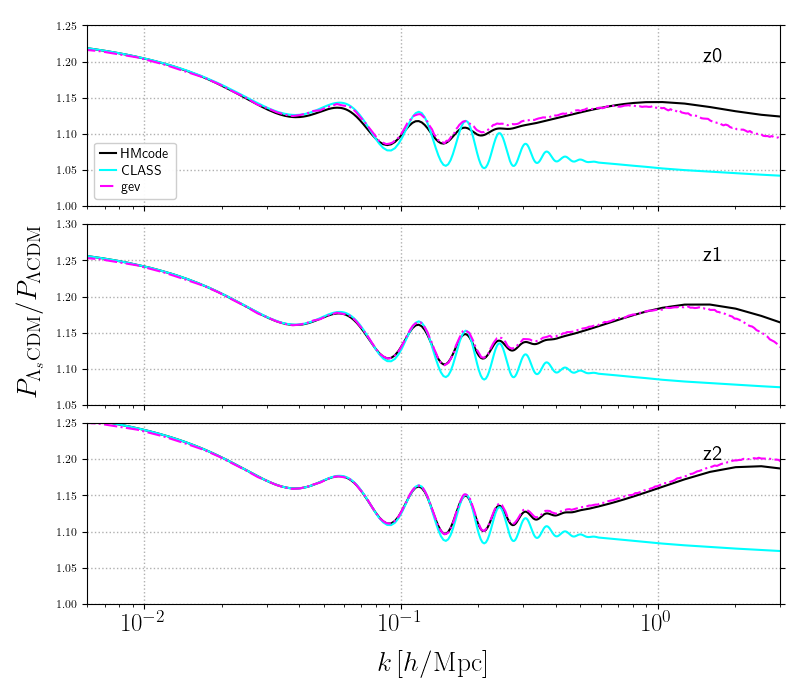}\hfil
  \includegraphics[width=0.453\textwidth]{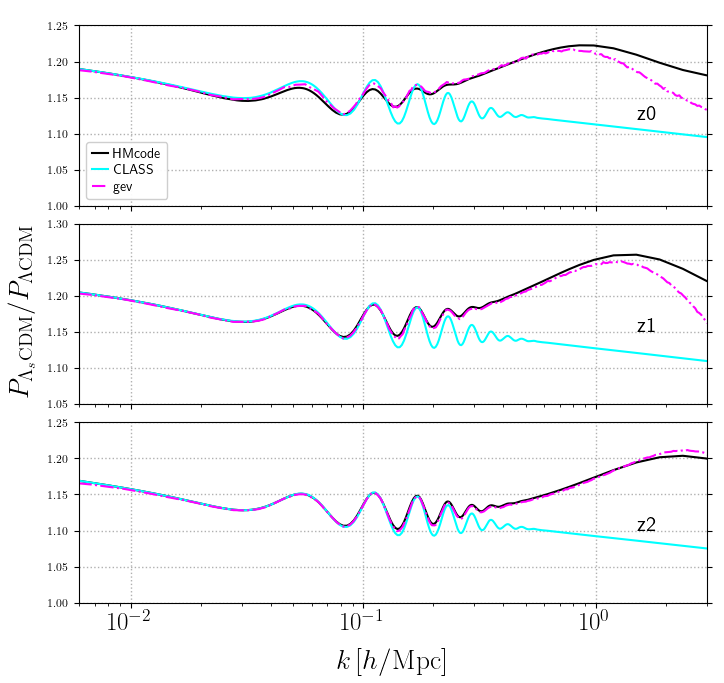}\\[-2pt]

  \parbox[t]{0.495\textwidth}{\centering (a) \textit{Planck}-only}\hfil
  \parbox[t]{0.495\textwidth}{\centering (b) full-dataset}\\

  \caption{$P_{\Lambda_{\rm s}{\rm CDM}}/P_{\Lambda{\rm CDM}}$ from \texttt{gevolution}, \texttt{CLASS}, and \texttt{HMCode} (implemented in \texttt{CLASS}) for (a) the \textit{Planck}-only and (b) the full-dataset best-fit parameters, at $z=2, 1,$ and $0$. The CDM-baryon (cb-only) curves from the N-body results are shown in Poisson gauge, whereas those from \texttt{CLASS} and \texttt{HMCode} are the synchronous-gauge spectra on linear scales (\texttt{CLASS} calculations are still performed in Newtonian gauge as in the main text). This does not affect our direct comparison at the redshifts of interest for $k\!\gtrsim\!0.02\,h\,\mathrm{Mpc}^{-1}$.}
  \label{model_ratios}
\end{figure*}

For completeness, we also consider nonlinear corrections to the CDM-baryon power spectrum predictions for all four cases studied in the manuscript using the \texttt{CLASS} implementation of \texttt{HMCode}~\cite{hmcode}. Fig.~\ref{model_ratios} shows the ratios $P_{\Lambda_{\rm s}{\rm CDM}}/P_{\Lambda{\rm CDM}}$ at $z=2,1$ and $0$ for both the \textit{Planck}-only and full-dataset best-fit parameters [panels (a) and (b), respectively]. Results from \texttt{HMCode} agree well with those from N-body simulations with \texttt{gevolution} on mildly nonlinear to nonlinear scales, with the remaining differences at the sub-percent level for all redshifts up to $k\simeq 1\,\kunit$. For $z=2$, the difference stays below $1\%$ throughout, whereas at $z=1$ and $0$ it reaches $\sim 3\%$ ($\sim 4\%$) around $k_{\max}=3\,\kunit$ for the \textit{Planck}-only (``full'') best-fit parameter set.

\begin{figure*}[t!]
  \centering

  \includegraphics[width=0.495\textwidth]{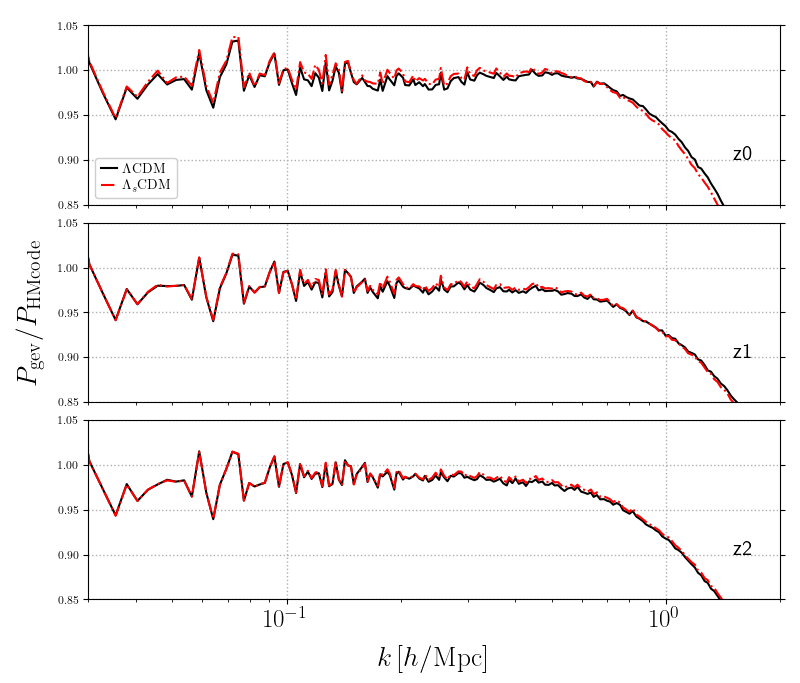}\hfil
  \includegraphics[width=0.495\textwidth]{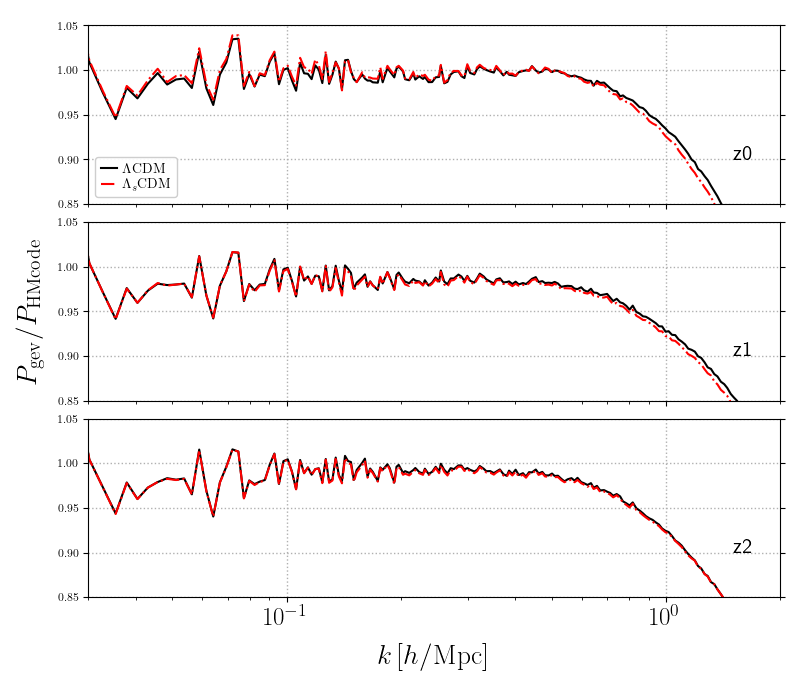}\\[-2pt]

  \parbox[t]{0.495\textwidth}{\centering (a) \textit{Planck}-only}\hfil
  \parbox[t]{0.495\textwidth}{\centering (b) full-dataset}\\

  \caption{Ratios of the absolute cb-only power spectra from N-body simulations with \texttt{gevolution} to \texttt{CLASS} predictions including nonlinear corrections from \texttt{HMCode}, for (a) the \textit{Planck}-only and (b) the full-dataset best-fit parameters, at $z=2, 1,$ and $0$.}
  \label{method_ratios}
\end{figure*}

In Fig.~\ref{method_ratios}, we switch to the \textit{absolute} spectra and investigate how the cb-only results from N-body simulations compare to \texttt{HMCode} for $\Lambda{\rm CDM}$ and $\Lambda_{\rm s}{\rm CDM}$ cosmologies, separately, again for the \textit{Planck}-only and full-dataset parameter sets [panels (a) and (b), respectively]. We observe strong consistency in how \texttt{HMCode} performs for the two models: down to the limit $k_{\rm f}=1.6\,h\,\mathrm{Mpc}^{-1}$ that we adopt for the absolute spectra, the $\Lambda{\rm CDM}$ and $\Lambda_{\rm s}{\rm CDM}$ curves agree within $0.5\%$ at $z=2$ and $1$ (only for the ``full'' dataset parameters this reaches $\sim 1\%$ at $z=1$). At $z=0$, the difference remains below $2\%$ for each parameter set.

\newpage

\bibliographystyle{apsrev4-2_mod}
\bibliography{biblio}

\end{document}